\documentclass[aps,prl,preprint,superscriptaddress]{revtex4-1}

\usepackage{color}
\usepackage{amsmath}
\usepackage{graphicx}
\usepackage{multirow}
\usepackage{amsfonts}
\usepackage{amssymb}
\usepackage{amscd}
\usepackage{multirow}
\usepackage{dcolumn}
\usepackage{bm}
\usepackage{float}

\newcommand{\bra}[1]{\mbox{$\left\langle #1 \right|$}}
\newcommand{\ket}[1]{\mbox{$\left| #1 \right\rangle$}}

\begin{document}

\title{Experimental measurement-device-independent quantum random number generation}

\author{You-Qi Nie}
\affiliation{Hefei National Laboratory for Physical Sciences at the Microscale and Department
of Modern Physics, University of Science and Technology of China, Hefei, Anhui 230026, China}
\affiliation{CAS Center for Excellence and Synergetic Innovation Center in Quantum Information
and Quantum Physics, University of Science and Technology of China, Hefei, Anhui 230026, China}

\author{Jian-Yu Guan}
\affiliation{Hefei National Laboratory for Physical Sciences at the Microscale and Department
of Modern Physics, University of Science and Technology of China, Hefei, Anhui 230026, China}
\affiliation{CAS Center for Excellence and Synergetic Innovation Center in Quantum Information
and Quantum Physics, University of Science and Technology of China, Hefei, Anhui 230026, China}

\author{Hongyi Zhou}
\affiliation{Center for Quantum Information, Institute for Interdisciplinary Information Sciences,
Tsinghua University, Beijing 100084, China}

\author{Qiang Zhang}
\affiliation{Hefei National Laboratory for Physical Sciences at the Microscale and Department
of Modern Physics, University of Science and Technology of China, Hefei, Anhui 230026, China}
\affiliation{CAS Center for Excellence and Synergetic Innovation Center in Quantum Information
and Quantum Physics, University of Science and Technology of China, Hefei, Anhui 230026, China}

\author{Xiongfeng Ma}
\email{xma@tsinghua.edu.cn}
\affiliation{Center for Quantum Information, Institute for Interdisciplinary Information Sciences,
Tsinghua University, Beijing 100084, China}

\author{Jun Zhang}
\email{zhangjun@ustc.edu.cn}
\affiliation{Hefei National Laboratory for Physical Sciences at the Microscale and Department
of Modern Physics, University of Science and Technology of China, Hefei, Anhui 230026, China}
\affiliation{CAS Center for Excellence and Synergetic Innovation Center in Quantum Information
and Quantum Physics, University of Science and Technology of China, Hefei, Anhui 230026, China}

\author{Jian-Wei Pan}
\affiliation{Hefei National Laboratory for Physical Sciences at the Microscale and Department
of Modern Physics, University of Science and Technology of China, Hefei, Anhui 230026, China}
\affiliation{CAS Center for Excellence and Synergetic Innovation Center in Quantum Information
and Quantum Physics, University of Science and Technology of China, Hefei, Anhui 230026, China}

\date{\today}

\begin{abstract}

The randomness from a quantum random number generator (QRNG) relies on the accurate
characterization of its devices. However, device imperfections and inaccurate characterizations can result in
wrong entropy estimation and bias in practice, which highly affects the genuine randomness generation and may even
induce the disappearance of quantum randomness in an extreme case.
Here we experimentally demonstrate a measurement-device-independent (MDI) QRNG
based on time-bin encoding to achieve certified quantum randomness even when the measurement devices are uncharacterized and untrusted.
The MDI-QRNG is randomly switched between the regular randomness generation mode and a test mode, in which four quantum states are randomly prepared to perform measurement
tomography in real-time.
With a clock rate of 25 MHz, the MDI-QRNG generates a final random bit rate of 5.7 Kbps.
Such implementation with an all-fiber setup provides an approach to construct a fully-integrated MDI-QRNG with trusted but error-prone devices in practice.
\end{abstract}


\maketitle 

Random numbers are widely required in a diversity of applications. Based on the fundamental laws of quantum physics, quantum random number generators (QRNGs) can produce true random numbers, which are unpredictable, irreproducible, and unbiased. So far, various QRNG schemes have been demonstrated including the ones based on beam splitters~\cite{Stefanov00,Jennewein00}, photon arrival times~\cite{Wayne10,Wahl11,Nie14}, vacuum fluctuations~\cite{Gabriel10,Symul11,Jofre11,STZ10,SCK16}, laser phase fluctuations~\cite{Qi10,Xu12,Nie15,Zhou15,AAM15,Nie16}, and time-frequency uncertainty~\cite{Xu16}. For a review of the subject, one can refer to~\cite{MaQRNG} and references therein.

A typical QRNG consists of two parts, randomness source and quantum measurement. For instance, in a simple prepare-and-measure scheme, the particles are prepared in a fixed quantum state, $\ket{+} = \frac{1}{\sqrt{2}}(\ket{0}+\ket{1})$, an eigenstate of the $X$ basis, and then they are measured in the $Z$ basis, so that the outcomes of `0' and `1' are produced with the equal probability as raw output data.
The central issue in QRNG is entropy estimation, i.e., how much genuine quantum randomness can be extracted from the raw data.
For each conventional QRNG implementation, all the devices have to be precisely characterized, and with properly modelling the min-entropy estimation is normally used to quantify the randomness of the output data~\cite{Ma13}. After randomness extraction, final random numbers can be obtained from the raw data.

In practice, the imperfections of realistic devices and inaccurate characterizations can result in wrong entropy estimation and bias to the output bits.
Such bias problem is very similar to the adversary scenario in quantum key distribution (QKD), where the intervention of an eavesdropper may introduce bias to the keys (from the adversary's point of view). Thus, in the data analysis of QRNG, one can introduce an adversary to model the bias problem. A QRNG can be regarded as a local machine packaged in a closed box,
and the imperfections may depend on some variables within the box, which the user is unaware of. The adversary may not have an access to control the variables directly. However, she has side information about how the variables evolve, which enables her to predict the working conditions of the devices and the outcome random numbers to some extent. Though the outcomes may still seem to be unbiased from the user's point of view, they are biased conditioned on the adversary's system. That is, the adversary might predict the outcomes partially. For some QRNG applications, especially the ones in cryptography, the drawback could cause security threats.

In order to effectively solve the problems of device imperfections and inaccurate characterizations, different QRNG protocols have been recently proposed to obtain certified genuine randomness even when devices are untrusted and uncharacterized~\cite{Pironio2010,Brunner15,Banik15,Ma15,Ma16,Xu16}, including device-independent QRNGs (DI-QRNGs) and semi-device-independent QRNGs. Not surprisingly, these concepts and techniques are all originated from QKD. The DI-QRNG protocol can produce certified randomness based on the violation of Bell¡¯s inequality~\cite{Pironio2010} without trusting the quantum devices. However, the DI-QRNG requires efficiency-loophole-free Bell tests, which makes the experimental implementation rather challenging and inefficient~\cite{Pironio2010}. In practice, there is a trade-off between system security and performance. By adding a few reasonable assumptions to the quantum devices, the DI-QRNG becomes much more practical~\cite{Banik15,Ma15,Ma16}, which is called semi-device-independent QRNG scheme. For instance, Lunghi \emph{et al.} have demonstrated a self-testing QRNG experiment with general device assumptions such as bounded dimensions without relying on detailed characterizations~\cite{Brunner15}. Cao \emph{et al.} have proposed and experimentally realized a source-independent QRNG based on entropic uncertainty relation of $X$ and $Z$ basis measurement given trusted measurement devices~\cite{Ma16}.

Similarly, one may ask an interesting question: whether it is possible to generate genuine quantum randomness when the measurement devices are uncharacterized and untrusted?
The answer to this question leads to the emergence of measurement-device-independent QRNG (MDI-QRNG) proposals~\cite{Banik15,Ma15}.
Considering the duality of state preparation and measurement and using the idea similar to source-independent QRNG scheme~\cite{Ma16}, given an assumption of trusted source part the MDI-QRNG is randomly switched between regular randomness generation mode and a test mode, in which different input states are randomly prepared to test the reliability of the measurement devices in real-time~\cite{Ma15}. In such a way, quantum random numbers can be generated even with uncharacterized and untrusted measurement devices.

\begin{figure} [bt]
\centering
\includegraphics[width=8.5 cm]{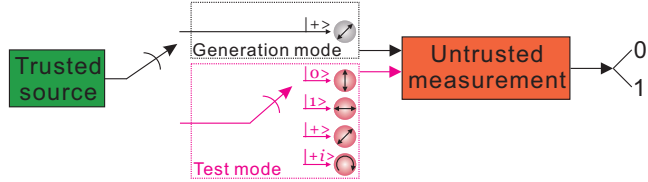}
\caption{Measurement-device-independent QRNG scheme.}
\label{fig1}%
\end{figure}

\emph{MDI-QRNG protocol.} The MDI-QRNG protocol is described in Fig~\ref{fig1}. The quantum states emitted from the trusted randomness source are measured by untrusted devices with a binary output `0' or `1'. In the generation mode, a fixed state $\ket{+}$ is sent. The user randomly chooses $N_0$ out of total $N$ turns as test mode, in which the source randomly emits quantum states $\ket{0}$, $\ket{1}$, $\ket{+}$ and $\ket{+i}$ ($\ket{+i} = \frac{1}{\sqrt{2}}(\ket{0}+i\ket{1})$) as test states to perform a measurement tomography. Here, $\ket{0}$, $\ket{1}$, $\ket{+}$ and $\ket{+i}$ are the eigenstates of Pauli matrices $\sigma_z$, $\sigma_z$, $\sigma_x$ and $\sigma_y$, respectively. In order to choose both the test mode and prepared test states, random number seeds are required. Therefore, it it crucial to guarantee that the randomness generation is larger than the randomness consumption.

The key idea of the scheme is self-testing, that is, it can be tested out whether the output random numbers are reliable according to the tomography results. We model the measurement using a qubit positive-operator valued measure (POVM)~\cite{PhysRevA.76.022325},
\begin{equation}
\begin{aligned}
F_0&=a_0(I+\vec{n}_0 \cdot \vec{\sigma}), \\
F_1&=a_1(I+\vec{n}_1 \cdot \vec{\sigma}),
\end{aligned}
\end{equation}
where $F_0$ and $F_1$ are the measurement outputs `0' and `1', respectively, $\vec{\sigma}=(\sigma_x, \sigma_y, \sigma_z)$ is the Pauli matrix vector, $\vec{n}_0=(n_x, n_y, n_z)$ and $\vec{n}_1$ are real number vectors. In the experiment, given the four test states, the probability of output `0' (`1') and the POVM parameters $a_0$, $n_x$, $n_y$, and $n_z$, can be evaluated, which are then used for randomness quantification of the raw data. The details are shown in Supplemental Material.

In the model of MDI-QRNG, the adversary can let her ancillary photons correlated with the photons emitted from the source, and she can perform a measurement on her ancillary photons to extract information about the output random numbers. We can classify the adversary into a classical one or a quantum one according to her ability.
Compared with a classical adversary who can only perform individual measurement on each ancillary photon,
a quantum adversary has the ability of performing joint measurement. That is, she can store her ancillary photons in a quantum memory and then a measurement is performed together~\cite{Yuan16}, which enlarges her side information compared with the classical scenario. In this paper, the randomness quantification is evaluated against a classical adversary, while the randomness quantification against a quantum adversary is different and deserves future work for clarification.

\begin{figure*}[hbt]
\centering
\includegraphics[width=18 cm]{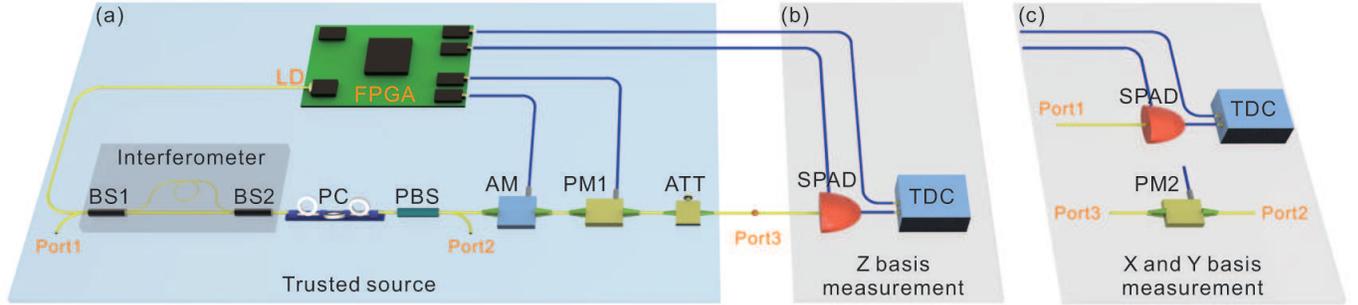}
\caption{Experimental setup of MDI-QRNG including the trusted source (a) and the $Z$ basis measurement part (b). During the verification process of prepared quantum states, the measurement part is reconfigured when $X$ and $Y$ basis measurement are performed (c). LD: laser diode, FPGA: field-programmable gate array, BS: beam splitter, PC: polarization controller, PBS: polarizing beam splitter, AM: amplitude modulator, PM: phase modulator, ATT: attenuator, SPAD: single-photon avalanche diode, TDC: time-to-digital converter.}
\label{fig2}%
\end{figure*}

\emph{Experimental setup.}
The time-bin encoding MDI-QRNG setup is shown in Fig.~\ref{fig2}.
Phase-randomized narrow optical pulses created from a 1550 nm laser diode (LD) with a clock rate of 25 MHz
are entered into an unbalanced interferometer with a time delay of 9.6 ns to form two time-bin pulses.
The output port of the interferometer is connected with a polarizing beam splitter (PBS) via a polarization controller (PC).
The PBS output is further modulated by two polarization-maintaining components, i.e.,
an amplitude modulator (AM) and a phase modulator (PM), which are controlled by a field-programmable gate array (FPGA),
to prepare four time-bin quantum states of $\ket{0}$, $\ket{1}$, $\ket{+}$ and $\ket{+i}$.

For $Z$ basis measurement as shown in Fig.~\ref{fig2}(b), photons emitted from the ATT output port (Port3) are detected by a fully integrated
1.25 GHz InGaAs/InP single-photon avalanche diode (SPAD) based on the technique of sine wave gating~\cite{Liang12}, with a detection efficiency of $\sim 25\%$.
The gate signals of the SPAD are synchronized with the laser pulses and the detection signals are further measured by a time-to-digital converter (TDC).
When $X$ ($Y$) basis measurement is required, the configuration in the measurement part is changed as Fig.~\ref{fig2}(c). Port3 is connected with Port2 via
an additional PM (PM2), so that emitted photons from the ATT reenter into the interferometer and photons at Port1 are finally detected by the SPAD. Such configuration
greatly simplifies the experimental setup without requiring additional unbalanced interferometer and auxiliary phase stabilization.

\begin{figure}[b]
\centering
\includegraphics[width=8.5cm]{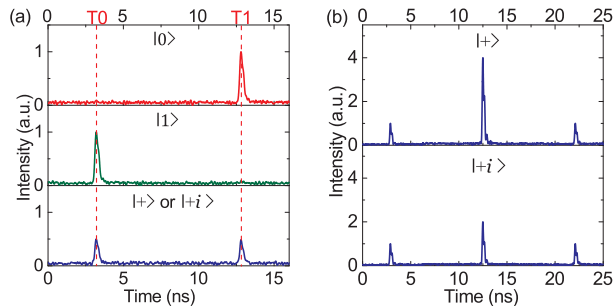}
\caption{Intensity traces of four prepared states of $\ket{0}$, $\ket{1}$, $\ket{+}$ and $\ket{+i}$ observed in an oscilloscope (a). The measurement part is configured as Fig.~\ref{fig2}(b) whilst the attenuator is set as the minimal value and a high-speed photodetector is used instead of the SPAD. Early and late time-bin pulses at $T0$ and $T1$ correspond to the states of $\ket{1}$ and $\ket{0}$, respectively. When both of the early and late pulses are attenuated by half and the phase of the early pulse is set as 0 ($\frac{\pi}{2}$) by PM1, the state of $\ket{+}$ ($\ket{+i}$) is prepared. However, the intensity traces in two cases are exactly the same.
To further distinguish the two states, the measurement part is configured as Fig.~\ref{fig2}(c). When the phase of PM2 is set as 0, $\ket{+}$ and $\ket{+i}$ produces constructive and intermediate interferences, respectively (b).}
\label{fig3}%
\end{figure}

Typical intensity traces of four time-bin states observed in an oscilloscope are plotted in Fig.~\ref{fig3}, in which a high-speed photodetector is used instead of the SPAD and
the attenuator is set as the minimal value. Given that a laser pulse is entered into the unbalanced interferometer, two time-bin pulses, i.e., an early pulse and a late pulse, are created. When the early (late) pulse is removed by the AM, the state of $\ket{0}$ ($\ket{1}$) is prepared, see the upper (middle) trace in Fig.~\ref{fig3}(a). When both of the pulses are attenuated by half due to the AM and meanwhile the relative phase of the early pulse is set as 0 ($\frac{\pi}{2}$) due to the PM1, the state of $\ket{+}$ ($\ket{+i}$) is prepared. However, from the intensity traces observed in the oscilloscope $\ket{+}$ and $\ket{+i}$ states cannot be distinguished, see the lower trace in Fig.~\ref{fig3}(a). To further distinguish $\ket{+}$ and $\ket{+i}$, the measurement part is configured as Fig.~\ref{fig2}(c),
and the phase of PM2 set as 0 ($\frac{\pi}{2}$) corresponds to $X$ ($Y$) basis measurement. Fig.~\ref{fig3}(b) shows the intensity difference in two cases when the phase of PM2 is 0, where the state of $\ket{+}$ ($\ket{+i}$) produces constructive (intermediate) interference.

\begin{figure}[tb]
\centering
\includegraphics[width=8.5cm]{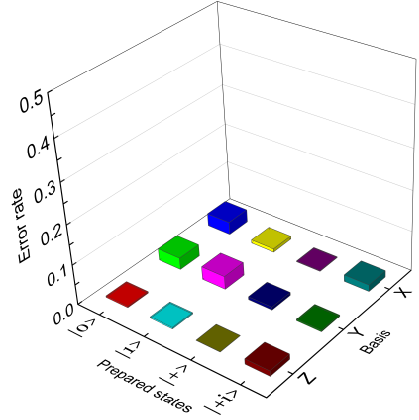}
\caption{The measured error rates of four prepared quantum states in the three orthogonal basis.}
\label{fig4}%
\end{figure}

To further verify the prepared states, the intensities of the time-bin pulses are attenuated to single-photon level via an attenuator (ATT) and the optimal mean photon number is set as $\sim$ 0.06 according to the theoretical model of MDI-QRNG~\cite{Ma15}. The time-bin states are then projected to $X$, $Y$ and $Z$ basis, respectively, and the measured results are shown in Fig.~\ref{fig4}, in which low error rates indicate the accuracy of the prepared quantum states. These error rates include the minor contributions due to the dark counts and afterpulses~\cite{ZIZ15} of the InGaAs/InP SPAD.

In the implementation process of the MDI-QRNG protocol, the source is operated either in generation mode or in test mode, whilst the measurement part is fixed at $Z$ basis.
In generation mode, fixed $\ket{+}$ state is sent and after the $Z$ basis measurement random bit `0' or `1' is generated.
In test mode, four states are randomly sent with equal probability to perform measurement tomography.
A large amount of random numbers are stored inside the FPGA in prior to determine which mode and which state to prepare for each laser pulse.
In order to finally gain output randomness higher than input randomness, the proportion of generation mode is much larger than that of test mode.

In each round of the MDI-QRNG process, $2^{34}$ quantum states in total including $2^{15}$ test states
are sent. For each test state, 34 random bits are used to determine its position in the sequence and further 2 bits are used to determine the state to be prepared.
The detection information (no-click, `0' or `1') of each quantum state is recorded.
Therefore, each round consumes 1152 Kbits of random numbers and produces 16 Gbits of raw data.
In the experiment, the MDI-QRNG process is performed for 100 rounds in total, so that around 115 Mbits of
random numbers are consumed and 1600 Gbits of raw data are produced.
The amount of prepared test states is $3.3 \times 10^{6}$ and their measurement tomography results are listed in Table~\ref{table1}.

\begin{table}
\centering
\caption{Results of measurement tomography.
\label{table1}}
\begin{tabular}{c|ccc}
  \hline
  Test state & Amount & Counts of `1' & Probability \\
  \hline
  $\ket{0}$ & 820318  & 121   & $1.48\times10^{-4}$   \\
  $\ket{1}$ & 818254  & 13067 & $1.60\times10^{-2}$   \\
  $\ket{+}$ & 819125  & 6431  & $7.85\times10^{-3}$   \\
  $\ket{+i}$ & 819103 & 6403  & $7.82\times10^{-3}$   \\
  \hline
\end{tabular}
\end{table}

Here we briefly introduce the randomness quantification, see Supplemental Material for details.
In the generation mode, based on the tomography results the lower bound of randomness against classical adversary is quantified with min-entropy~\cite{Ma15}
\begin{equation}\label{eq:randomnessmain}
R(F_0,F_1)\geq 2a_0H_{\infty}\left(\frac{1+\sqrt{1-n_y^2-n_z^2}}{2}\right),
\end{equation}
which is also suited for high-dimensional POVMs and the scenario that the adversary performs different POVMs for different turns. A practical MDI-QRNG system suffers two main problems, statistical fluctuation and imperfect qubit source. Note that other experimental imperfections such as transmission loss are also included in the form of POVM. We simply follow Ref.~\cite{Ma15} to address these two issues.

The number of total turns is finite and the statistical fluctuations should be taken into consideration, i.e., the measurement tomography may not be accurate due to the finite data effect. Given the test state $\rho_i$ $(i=1,2,3,4)$, i.e., $\rho_1=\ket{0}\bra{0}$, $\rho_2=\ket{1}\bra{1}$, $\rho_3=\ket{+}\bra{+}$ and $\rho_4=\ket{+i}\bra{+i}$, let $N_i$ be the number of test turns, $N_0$ be the number of generation turns, and $p_i$ ($p'_i$) be the probability of output `0' in test (generation) turns. The key point of the statistical fluctuation analysis is to use $p_i$ (measured value) to bound the parameter $p'_i$. When the data size is large enough, $p'_i\approx p_i$. In the finite data case, there is a deviation between $p_i$ and $p'_i$, denoted by $\theta_i$. Given the number of turns $N_i$ ($i=0,1,2,3,4$), $\theta_i$ is a function of $p_i$.

In the experiment, a weak coherent state source is used, which is, however, an imperfect qubit source. Given a coherent state source with an intensity of $\mu$, after phase randomization, it becomes a mixture of photon number states following a Poisson distribution. Such imperfection would affect the final randomness evaluation by~\cite{Ma15}
\begin{equation}\label{eq:randomnessforcoherentsource}
R(F_0,F_1)\geq \min_{a_0,n_y,n_z} \frac{2a_0(1+\mu)}{e^\mu}H\left(\frac{1+\sqrt{1-n_y^2-n_z^2}}{2}\right),
\end{equation}
with constraints $|n_x|^2+|n_y|^2+|n_z|^2=1$, $0\le a_0\le1$, and
\begin{equation}
\label{eq:constraint}
\begin{aligned}
(a_0+a_0n_z)(1+\mu)e^{-\mu} & \leq p_1 \pm \theta_1  \leq (a_0+a_0n_z)(1+\mu)e^{-\mu} \\
                            & +1-e^{\mu}-\mu e^{-\mu},\\
(a_0-a_0n_z)(1+\mu)e^{-\mu} & \leq p_2 \pm \theta_2  \leq (a_0-a_0n_z)(1+\mu)e^{-\mu} \\
                            & +1-e^{\mu}-\mu e^{-\mu},\\
(a_0+a_0n_x)(1+\mu)e^{-\mu} & \leq p_3 \pm \theta_3  \leq (a_0+a_0n_x)(1+\mu)e^{-\mu} \\
                            & +1-e^{\mu}-\mu e^{-\mu},\\
(a_0+a_0n_y)(1+\mu)e^{-\mu} & \leq p_4 \pm \theta_4  \leq (a_0+a_0n_y)(1+\mu)e^{-\mu} \\
                            & +1-e^{\mu}-\mu e^{-\mu}. \\
\end{aligned}
\end{equation}
We make a worst-case assumption that the multi-photon components can be fully manipulated by the untrusted measurement devices and thus cannot generate output randomness. In such a way, the source intensity $\mu$ can be optimized for the output randomness from Eq.~\eqref{eq:randomnessforcoherentsource}.

\begin{figure}[t]
\centering
\includegraphics[width=8.5 cm]{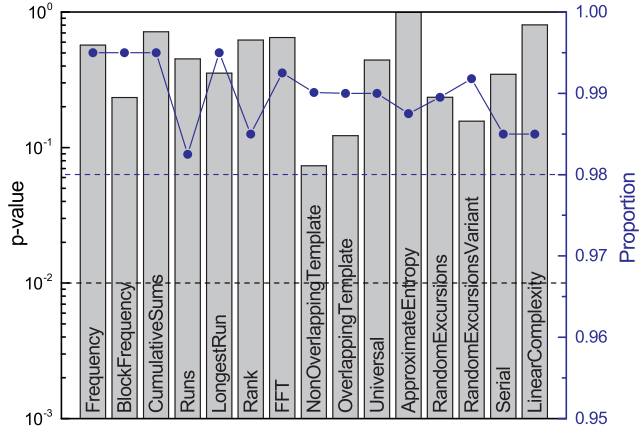}
\caption{
The NIST test results of the final random data with a size of 390 Mbits. Given an item, when the $p$-value (column) and the proportion (dot) are more than 0.01 and 0.98, respectively, it means that the random data pass the item.}
\label{fig5}%
\end{figure}

Employing the analysis method shown in Eq.~\eqref{eq:randomnessforcoherentsource} and Eq.~\eqref{eq:constraint} for the experimental results shown in Table~\ref{table1}, the min-entropy of the MDI-QRNG is lower-bounded by $2.3\times10^{-4}$ bit per pulse. For randomness extraction, a Toeplitz-matrix hash function is applied. The final random number generation rate is 5.7 Kbps. We finally obtain 390 Mbit random numbers, which are 3.4 times larger than the amount of randomness consumed as seeds. In order to verify the quality of the final random bits, the standard NIST statistical tests are applied~\cite{NIST}. Clearly, the final random bits pass all the test items as shown in Fig.~\ref{fig5}.


In summary, we experimentally realize a practical measurement-device-independent quantum random number generator using time-bin encoding.
The output randomness against classical adversary can be certified and quantified, even when the measurement devices are uncharacterized and untrusted.
After randomness quantification, the min-entropy of MDI-QRNG reaches $2.3\times10^{-4}$ bits per pulse, corresponding to final random number generation rate of 5.7 Kbps. Moreover, the ratio of random number generation to random number consumption is 3.4.
This all-fiber experimental setup exhibits the feasibility of constructing a fully-integrated and compact MDI-QRNG with trusted but error-prone devices in practice.

\begin{acknowledgments}
This work has been financially supported by the National Basic Research Program of China Grant No.~2013CB336800, the National Natural Science Foundation of China Grant No.~61275121, and the Chinese Academy of Sciences. Y.-Q. Nie, J.-Y. Guan and H. Zhou contributed equally to this work.
\end{acknowledgments}


\section{Supplemental Material: Experimental measurement-device-independent quantum random number generation}

\subsection{Measurement tomography}
The security analysis of the experiment is mainly based on the previous theoretical work~\cite{Ma15}.
In this section, we summarize the key step of the analysis: measurement tomography.

The user prepares (trusted) qubit states for the (untrusted) measurement device, which outputs bits `0' and `1'. In practice, if the device outputs other signals, such as losses and double clicks, the user directly assigns the bit '0'. From the user's point of view, the measurement is a qubit POVM though the actual POVM may be high-dimensional. In fact, it can be proved that in order to let the user obtain the least randomness, the best strategy for an adversary is a qubit POVM~\cite{Ma15}. An arbitrary qubit POVM can be expressed as
\begin{equation}
\begin{aligned}
F_0&=a_0(I+\vec{n}_0 \cdot \vec{\sigma})\\
F_1&=a_1(I+\vec{n}_1 \cdot \vec{\sigma}),
\end{aligned}
\end{equation}
where the coefficients $a_0$ and $a_1$ are real numbers and $\vec{n_0}$ and $\vec{n_1}$ are real vectors satisfying
\begin{equation}
\begin{aligned}
a_0,a_1&\geq0 \\
a_0+a_1&=1 \\
|n_0|,|n_1|&\leq 1 \\
a_0\vec{n}_0+a_1\vec{n}_1&=0.
\end{aligned}
\end{equation}
The probabilities of output bits `0' and `1' given an input state $\rho$ are
\begin{equation}\begin{aligned}
P(0|\rho)&=tr(F_0\rho)\\
P(1|\rho)&=tr(F_1\rho).
\end{aligned}
\end{equation}
When the input states are $\ket{0}$, $\ket{1}$, $\ket{+}$ and $\ket{+i}$, the corresponding probabilities of output bit `0' are
\begin{equation}
\begin{aligned}\label{eq:tomography}
  p_1=P(0\big|\ket{0}\bra{0})&=a_0+a_0n_z \\
  p_2=P(0\big|\ket{1}\bra{1})&=a_0-a_0n_z \\
  p_3=P(0\big|\ket{+}\bra{+})&=a_0+a_0n_x \\
  p_4=P(0\big|\ket{+i}\bra{+i})&=a_0+a_0n_y,
\end{aligned}
\end{equation}
where $p_i$ $(i=1,2,3,4)$ can be estimated in the experiment.
From Eq.~\eqref{eq:tomography}, one can find that the number of unknown parameters is equal to that of equations.
Therefore, the POVM parameters $a_0$, $n_x$, $n_y$, and $n_z$ can be calculated and the measurement tomography is accomplished.

\subsection{Data analysis}
For a qubit source, the lower bound of randomness against a classical adversary is given by~\cite{Ma15}
\begin{equation}\label{app:randomnessmain}
R(F_0,F_1)\geq 2a_0H_{\infty}\left(\frac{1+\sqrt{1-n_y^2-n_z^2}}{2}\right),
\end{equation}
which is for the asymptotic case with infinite data size.

In the experiment, the data size is finite, and hence statistical fluctuations should be taken into consideration. That is, the measurement tomography may not be accurate due to the finite key effect. Let $N_i$ be the number of turns with input state $\rho_i$ $(i=1,2,3,4)$ in the test turns and $N_0$ be the number of generation turns, where $\rho_1=\ket{0}\bra{0}$, $\rho_2=\ket{1}\bra{1}$, $\rho_3=\ket{+}\bra{+}$ and $\rho_4=\ket{+i}\bra{+i}$.
Let $p_i$ ($p'_i$) be the conditional probability of output `0' in test (generation) turns given input states $\rho_i$. $p_i$ can be used to estimate $p'_i$. In the case of finite data, there is a deviation $\theta_i$ between $p_i$ and $p'_i$, which decreases as the data size increases. Similar to the phase error estimation in quantum key distribution (QKD), the failure probability $\epsilon_{\theta}$ in our estimation is given by \cite{Ma2011Finite}
\begin{equation}\label{eq:statisticalfluctuations}
\epsilon_{\theta}=P(p'_i>p_i+\theta_i)\leq \frac{4\sqrt{N_i+N_0}}{\sqrt{N_iN_0(1+p_i)(1-p_i)}}2^{-(N_i+N_0)\xi_i(\theta_i)},
\end{equation}
where $\xi_i(\theta_i)$ is defined as
\begin{equation}
\xi_i(\theta_i)=H(\frac{1+p_i}{2}+\frac{N_0\theta_i}{N_0+N_i})-\frac{N_iH((1+p_i)/2)+N_0H((1+p_i)/2+\theta_i)}{N_0+N_i}.
\end{equation}
The statistical fluctuation $\theta_i$ can be calculated given the values of $N_i$ and $\epsilon_{\theta}$. In the worst case, $a_0$, $n_y$ and $n_z$ should take the lower bound values,
\begin{equation}
\begin{aligned}
n_y&=\frac{2p_4-2\theta_4}{p_1+p_2+\theta_1+\theta_2}-1,\\
n_z&=\frac{2p_1-2\theta_1}{p_1+p_2+\theta_1+\theta_2}-1,\\
a_0&=\frac{p_1+p_2-\theta_1-\theta_2}{2}.
\end{aligned}
\end{equation}
Then, the parameters $n_y$,$n_z$ and $a_0$ are substituted into Eq.~\eqref{app:randomnessmain} to calculate the randomness lower bound.

In the experiment, a coherent state source rather than a qubit source is used. The photon number follows a Poisson distribution. After phase randomization, the state becomes a Fock state mixture. The source contains three kind of components, i.e., vacuum, single photon and multi-photon. In such case, the corresponding randomness is lower bounded by
\begin{equation}\label{app:randomnessforcoherentsource}
R(F_0,F_1)\geq \min_{a_0,n_y,n_z} \frac{2a_0(1+\mu)}{e^\mu}H\left(\frac{1+\sqrt{1-n_y^2-n_z^2}}{2}\right),
\end{equation}
where the minimization of $a_0$, $n_y$ and $n_z$ is due to the fact that statistical fluctuations are considered. These parameters cannot be obtained directly from the experiment. Instead, they can be estimated.

Here, we take the worst-case assumption that the multi-photon (bad) component cannot generate any randomness. Also, we follow the argument that the vacuum and single photon (good) part can be regarded as an effective single photon when performing the tomography~\cite{Ma15}. The probability of output bit `0' given the four input test states $\rho_i$ are
\begin{equation}
p(0|\rho_i)=(1+\mu)e^{-\mu}P_i+(1-e^{\mu}-\mu e^{-\mu})P'_i,
\end{equation}
where $P_i$ and $P'_i$ are the probabilities of output bit `0' when the input state comes from the good and bad parts, respectively. By letting $P'_i$ be `0' or `1', we can derive the constraints,
\begin{equation}\label{eq:constraints}
\begin{aligned}
   (a_0+a_0n_z)(1+\mu)e^{-\mu}&\leq p_1 \leq (a_0+a_0n_z)(1+\mu)e^{-\mu}+1-e^{\mu}-\mu e^{-\mu}, \\
   (a_0-a_0n_z)(1+\mu)e^{-\mu}&\leq p_2 \leq (a_0-a_0n_z)(1+\mu)e^{-\mu}+1-e^{\mu}-\mu e^{-\mu}, \\
   (a_0+a_0n_x)(1+\mu)e^{-\mu}&\leq p_3 \leq (a_0+a_0n_x)(1+\mu)e^{-\mu}+1-e^{\mu}-\mu e^{-\mu}, \\
   (a_0+a_0n_y)(1+\mu)e^{-\mu}&\leq p_4 \leq (a_0+a_0n_y)(1+\mu)e^{-\mu}+1-e^{\mu}-\mu e^{-\mu}.\\
 \end{aligned}
\end{equation}

Further, we take account of statistical fluctuations in Eq.~\eqref{eq:constraints}. Given the failure probability $\epsilon_\theta$ and the number of turns $N_i$ ($i=0,1,2,3,4$), $\theta_i$ can be calculated by letting the inequality in Eq.~\eqref{eq:statisticalfluctuations} be an equation. Then, we have the following constraints,
\begin{equation}\label{eq:constraintswithfluc}
\begin{aligned}
   (a_0+a_0n_z)(1+\mu)e^{-\mu}&\leq p_1 \pm \theta_1 \leq (a_0+a_0n_z)(1+\mu)e^{-\mu}+1-e^{\mu}-\mu e^{-\mu} \\
   (a_0-a_0n_z)(1+\mu)e^{-\mu}&\leq p_2 \pm \theta_2 \leq (a_0-a_0n_z)(1+\mu)e^{-\mu}+1-e^{\mu}-\mu e^{-\mu} \\
  (a_0+a_0n_x)(1+\mu)e^{-\mu}&\leq p_3 \pm \theta_3 \leq (a_0+a_0n_x)(1+\mu)e^{-\mu}+1-e^{\mu}-\mu e^{-\mu} \\
   (a_0+a_0n_y)(1+\mu)e^{-\mu}&\leq p_4 \pm \theta_4 \leq (a_0+a_0n_y)(1+\mu)e^{-\mu}+1-e^{\mu}-\mu e^{-\mu}.\\
 \end{aligned}
\end{equation}
In the data post-processing, one needs to numerically solve the minimization problem of Eq.~\eqref{app:randomnessforcoherentsource} with the constraints in Eq.~\eqref{eq:constraintswithfluc}. Also, from Eq.~\eqref{app:randomnessforcoherentsource} one can optimize the source intensity $\mu$. In the experiment, the optimal value is around $\mu=0.06$.

\end{document}